\documentclass[12pt]{iopart} % provides new 2008 pss two-column style (no alternative manuscript style output available at present)
\usepackage{amssymb,graphicx}

\begin{document}

% Title of the article
\title[Ga$^+$-irradiation influence in FIB]{The influence of Ga$^+$-irradiation on the transport properties of
mesoscopic conducting thin films }

% Abbreviated title for the page headers
%\titlerunning{Transport Properties of Nanostructures}

\author{J. Barzola-Quiquia$^1$, S. Dusari, G. Bridoux, F. Bern, A. Molle and
P. Esquinazi$^2$}
\address{Division of Superconductivity and
Magnetism, Universit\"{a}t Leipzig, Linn\'{e}stra{\ss}e 5, D-04103
Leipzig, Germany}

\ead{$^1$j.barzola@physik.uni-leipzig.de;$^2$esquin@physik.uni-leipzig.de}

\begin{abstract}
We studied the influence of 30~keV Ga$^+$-ions -- commonly used in
focused ion beam (FIB) devices -- on the transport properties of thin
crystalline graphite flake, La$_{0.7}$Ca$_{0.3}$MnO$_3$ and Co thin
films. The changes of the electrical resistance were measured in-situ
during irradiation and also the temperature and magnetic field
dependence before and after irradiation.  Our results show that the
transport properties of these materials strongly change at Ga$^+$
fluences much below those used for patterning and ion beam induced
deposition (IBID), limiting seriously the use of FIB when the
intrinsic properties of the materials of interest are of importance.
We present a method that can be used to protect the sample as well as
to produce selectively irradiation-induced changes.
\end{abstract}

\submitto{\NT}
\maketitle

\section{Introduction}

In the last years a focused ion beam (FIB) of Ga$^+$-ions for etching
\cite{mat91} has attracted the attention of the community as an
alternative and flexible method to produce micro- and nanostructures
of materials, especially where the use of conventional methods
appears to be limited. This FIB technique has been successfully used
for nanostructuring different materials like magnetic and
superconducting \cite{gie05,tru,tse05} or more recently to study the
conduction behavior in metallic constrictions \cite{fer08}. One of
the advantages of this technique is its versatility; the use of any
resist appears, a priory, unnecessary. Nowadays, FIB devices are also
used for deposition of metallic or insulating materials with the help
of the same Ga$^+$-ions. These ions induce a decomposition of a
chemical metal precursor over the surface in question, a technique
called ion-beam induced or assisted deposition (IBID,IBAD)
\cite{mel87,mat96,lan02}. The main advantage of IBID/EBID is the
deposition of the desired patterns of a material without the need of
a mask or a pre-structured pattern using optical or e-beam
lithography (EBL). Also the possibility to modify only parts of the
patterns in electronic devices within nanometer dimensions is other
of the FIB advantages.

The modification of different properties of different materials
 has been studied in the past, as e.g. in
 magnetic La$_{0.7}$Sr$_{0.3}$MnO$_3$ thin films \cite{pal08} or
 Co/Pt multilayers  \cite{hyn01}. However, a fundamental
 problem of FIB was less discussed in
literature, namely the modification of the sample near surface region
and to a certain extent also its interior and their influence in the
transport properties by the use of Ga$^+$ions of energies up to
30~keV and fluences below $10^{12}$~cm$^{-2}$ in usual devices. We
note that before cutting or depositing material, the use of FIB
requires the precise alignment of the Ga-beam and this is done taking
a picture of the region in question irradiating it with the same
Ga$^+$-ions. Depending on the surface properties of the material in
question, in general Ga$^+$ fluences $\gtrsim 10^{11}~$cm$^{-2}$ are
used.  As we demonstrate below, these Ga$^+$-fluences necessary for
the first alignment may already affect seriously the intrinsic
properties of the material of interest  and can lead to wrong
interpretations of the effects that a reduction of the sample
geometry may produce.

The influence of Ga irradiation during the FIB preparation processes
was not yet quantitatively studied, neither in situ nor after
irradiation for fluences below $10^{12}$~cm$^{-2}$.  Specially when
materials are selected to investigate their transport properties
while their size is being reduced, care should be taken since the
electrical transport can be sensitive to lattice defects as well as
to the produced Ga contamination. The aim of this paper  is to report
on the changes in the electrical resistivity measured in-situ and its
temperature $T$ and magnetic field $B$ dependence after irradiation
of three different thin film materials to show the influence of the
Ga irradiation used in FIB devices and at fluences as low as the ones
used for beam alignment. The experiments were realized in two stages.
In the first stage we have done in-situ measurements of the
electrical resistance during irradiation and in the second stage its
resistance was measured as  a function of $T$ and $B$ to compare with
the corresponding virgin states. Using graphite as a test material
because its electrical resistance is extremely sensitive to lattice
defects \cite{arn09}, we provide in this work a possible solution
that can be used to strongly diminish the effects due to the
Ga$^+$-irradiation on different materials.

\section{Experimental Details} \label{details}

\begin{table}[b]
  \caption{Samples used, their dimensions (total
  length $\times$ width  $\times$ thickness) and the Ga$^+$-fluence
  irradiated on the samples. The fluence number in brackets refers to the total
  fluence after the second irradiation. The corresponding data of the Co-sample \#2 is
  given in the Table. The Co\#1 sample  had a thickness of
  57~nm. Other dimensions can be taken from Fig.~\protect\ref{samples_2}(c).}
  \begin{tabular}[htbp]{@{}llll@{}}
    \hline
    Sample & graphite  & La$_{0.7}$Ca$_{0.3}$MnO$_3$ & Co\#2 \\
    \hline
    Dimensions ($\mu$m)& $11\times 2 \times 0.015$ & $52 \times 7.3 \times 0.035$ & $18 \times 0.9 \times 0.022$ \\
    Fluence ($10^{11}$cm$^{-2}$) & 5 (10)  & $\geqslant 2.2$ & $\geqslant 2.2$ \\
\hline
  \end{tabular}
  \label{table}
\end{table}

\begin{figure}[htb]%
\includegraphics[width=\textwidth]{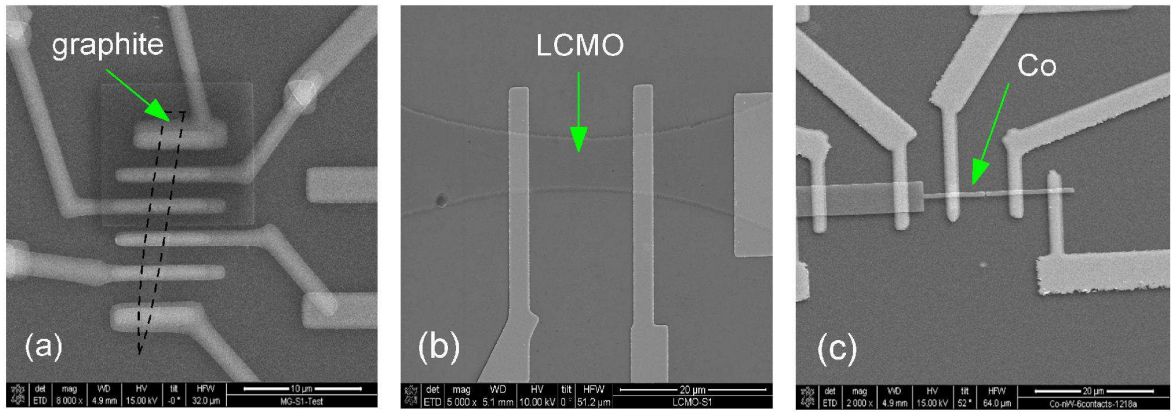}
\caption{Scanning electron microscope images of: (a) the graphite
flake (dashed line denotes its borders) with the six Au/Pt contacts.
(b) The LCMO film with the two inner voltages electrodes and one of
the input current electrodes. (c) The Co microwire \#1 with
electrodes at different positions. The irradiation has been made in
the whole region and the electrical resistance was measured between
the third and second electrodes from right.} \label{samples_2}
\end{figure}

We have used the FIB capabilities of  a FEI NanoLab XT 200 Dual Beam
microscope (DBM). The acceleration voltage was fixed in all our
experiments to 30~kV. The ion current and the area to be irradiated
were changed in order to obtain different Ga$^+$-fluences, see
Table~\ref{table}.  The selected samples were a crystalline graphite
flake prepared by a rubbing and ultrasonic process and
pre-characterized with electron backscattering diffraction (EBSD) and
Raman scattering \cite{bar08,arn09},  a La$_{0.7}$Ca$_{0.3}$MnO$_3$
(LCMO) thin film prepared by plasma laser deposition (PLD) and
micro-structured by EBL and wet etching process \cite{barpat}, and
thermal evaporated Co thin films (\#1 and \#2) previously structured
by EBL, see Fig.~\ref{samples_2}(a-c) and Table~\ref{table}.

Low-noise four-wires resistance measurements  (two for the input
current and two for the voltage measurement, important to eliminates
contributions of the lead resistance) have been performed with the AC
technique (Linear Research LR-700 Bridge with 8 channels LR-720
multiplexer) with ppm resolution and in some cases also with the DC
technique (Keithley 2182 with 2001 Nanovoltmeter and Keithley 6221
current source).

The Au/Pt lead contacts used for all samples were prepared by EBL
process using a e-beam resist PMMA 950K of $\sim 200~$nm thickness.
The lithography process was done with the Raith ELPHY Plus system
included in our microscope. The Au/Pt deposition of the contact
electrodes was done by evaporation in a high-vacuum chamber with a
nominal thickness of 25 and 9~nm, respectively. The in-situ
resistance measurements performed before and during the irradiation
of the sample  were done using a self-made sample holder fixed inside
the microscope chamber. The temperature and magnetic field dependence
measurements were performed using a commercial cryostat with a
temperature stability of 0.1~mK at 100~K. The magnetic field
generated by a superconducting solenoid was always applied normal to
the sample and input current.

To avoid or diminish irradiation effects we protected the graphite
sample with a negative electron beam resist (AR-N 7500) of thickness
of 300~nm.  In order to test the effectiveness of the resist film to
avoid contamination during irradiation, part of the graphite flake
was covered with the above mentioned resist in an additional process
after the Au/Pt leads were deposited on the sample, see
Fig.~\ref{samples_2}(a). This resin protects  the graphite sample in
the region of the three upper electrodes, allowing us to compare the
change in voltage in the unprotected, protected and intermediate
regions.

The advantages of this resist is that it allows us the patterning
by EBL in the desired shape, it is sufficiently robust in the used
 temperature range and it is a very bad electrical
conductor. The penetration depth of the Ga$^+$-ions in this resist as
well as in the samples and the distribution of the density of
vacancies as a function of sample depth were estimated taking into
account their density and using Monte Carlo simulations given by the
stopping and range of ions in matter (SRIM) \cite{ziegler,Ziegler2}
taking into account the energy of the Ga$^+$-ions, see
Fig.~\ref{SRIM}.

\section{Results and discussion}\label{results}

\subsection{In-situ transport measurements}

\begin{figure}[htb]%
\includegraphics[width=0.9\textwidth]{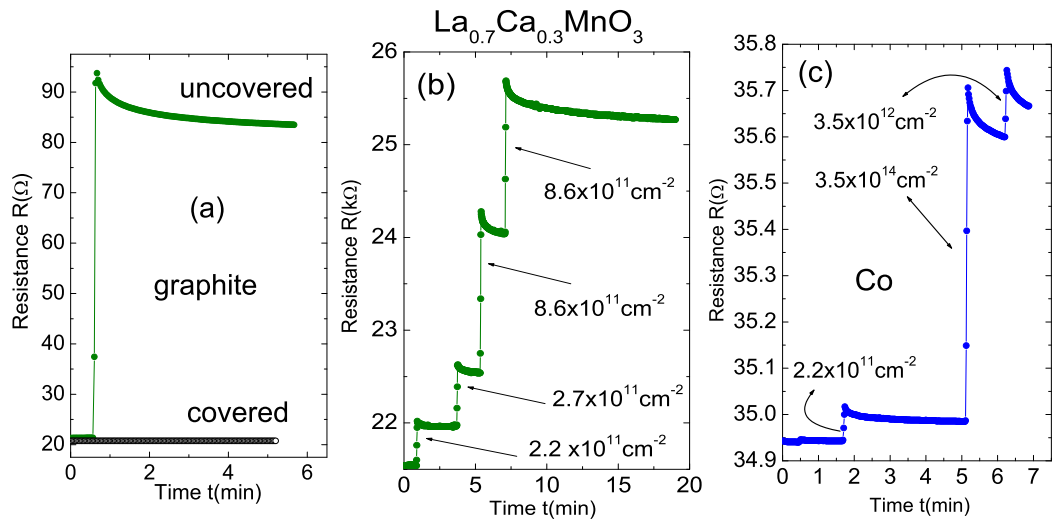}
\caption{Resistance as a function of time before, during and after
Ga$^+$ irradiation inside the FIB chamber for the samples (a)
crystalline graphite flake (first irradiation with a fluence of $5
\times 10^{11}$~cm$^{-2}$, (b) LCMO film and (c) Co film \#1. For
these two samples the used fluences are written in the figures.
All these measurements were done in-situ and at room temperature.
} \label{rgra}
\end{figure}

\begin{figure}[htb]%
\includegraphics[width=0.6\textwidth]{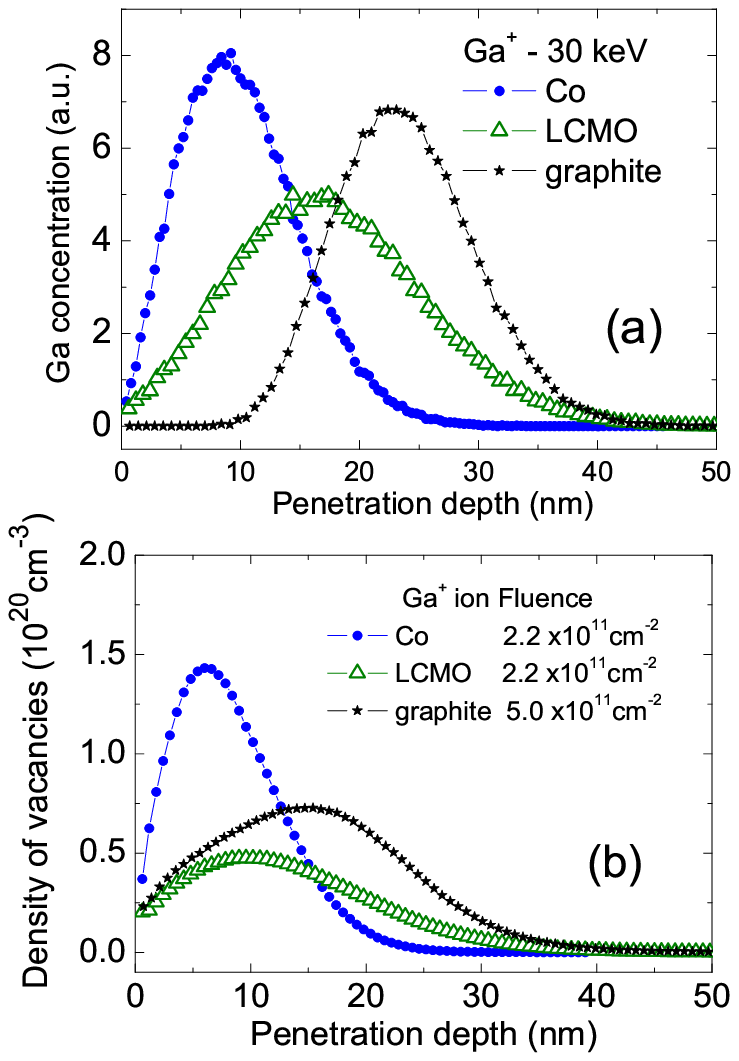}
\caption{(a) Qualitative distribution of implanted Ga ions as a
function of penetration depth into the sample for the three studied
materials. (b) Similarly for the vacancy density but taking into
account fixed fluences. The curves were obtained using the stopping
and range of ions in matter SRIM or IIS
\protect\cite{ziegler,Ziegler2} taking into account the energy of the
Ga$^+$-ions and the material densities.} \label{SRIM}
\end{figure}

A detailed  study of the electrical resistance of the above mentioned
materials was realized  {\em in-situ} during the Ga$^+$-ion
irradiation in the microscope chamber. In the case of the graphite
sample we measured simultaneously the resistance of the covered and
uncovered parts before, during and after irradiation.
Figure~\ref{rgra} shows the changes observed in this sample during
and after the first irradiation of  fluence  $5 \times 10
^{11}~$cm$^{-2}$. This fluence produces nominally $10^3$~ppm vacancy
concentration inside the sample whereas the Ga concentration
implanted is less than 1~ppm, see Fig.~\ref{SRIM}. The disorder
produced by the irradiation increases the resistance of the uncovered
part of this sample by a factor $> 4$.

The resistance of the covered part remains unchanged within $10^{-4}$
relative change indicating that the 300~nm thick resist  was enough
to stop the Ga$^+$-ions as expected since according the SRIM
calculations the maximal penetration of the Ga$^+$-ions in the resist
should be $\simeq 75~$nm. Immediately after irradiation the
resistance of the uncovered part starts to decay exponentially with
two characteristic relaxation times, as has been also observed after
proton irradiation at room temperature \cite{arn09}. This time
relaxation is observed for all samples just after the irradiation
finishes. This decay is related to local thermal relaxation process
and  to the diffusion of carbon interstitials and vacancies
\cite{niw95,lee05_vac}.

Qualitatively speaking, similar resistance changes during the
irradiation process are also observed in the two other samples, see
Fig.~\ref{rgra}(b,c). In the case of the LCMO sample the maximal
penetration of the ions is $\sim 45~$nm, see Fig.~\ref{SRIM}(a).
Because the thickness of this sample is 35~nm, part of the
Ga$^+$-ions are expected to be implanted and the rest to go through
the sample generating a considerable amount of defects as can be seen
in the calculated curves, see Fig.~\ref{SRIM}(b). In the case of the
Co wire \#1 the maximal ion penetration is $\sim 30~$nm~$ < 57~$nm
thickness, see Fig.~\ref{SRIM}, therefore the produced defects plus
Ga implantation are responsible for the relatively small increase in
the electrical resistance, see Fig.~\ref{rgra}(c). To study the
influence on the transport properties mainly due to the produced
defects by the Ga$^+$ irradiation, a second Co~\#2 sample with less
thickness has been studied and its results are discussed below.

The effect of the Ga irradiation to sample volume expansion
(thickness swelling) as well as milling (thickness reduction) depends
on the ion fluence, ion energy and target material. As an example we
refer to the work done in Ref.~\cite{pal08} where such studies were
done on La$_{0.7}$Sr$_{0.3}$MnO$_3$ thin films. According to this
work and taking into account our used fluences, any thickness
increase as well as any milling are completely negligible and do not
affect the changes measured in the resistance.

\subsection{Temperature and magnetic field dependence before and after irradiation}

1.{\em Graphite}: The special lattice structure of graphite and
the weak coupling between graphene layers make graphite a quasi-2D
semimetal, which carrier density depends strongly on the lattice
defects like vacancies and/or impurities. In a recent work it was
demonstrated that the electrical resistance of thin graphite
crystals of micrometer size changes after inducing less than 1~ppm
vacancy concentration by ion irradiation \cite{arn09}. This makes
graphite a extraordinary sensor for testing the efficiency of the
resin cover as well as to show the dramatic changes produced by a
relatively weak Ga$^+$-ion irradiation fluence.

\begin{figure}[htb]%
\includegraphics[width=0.50\textwidth]{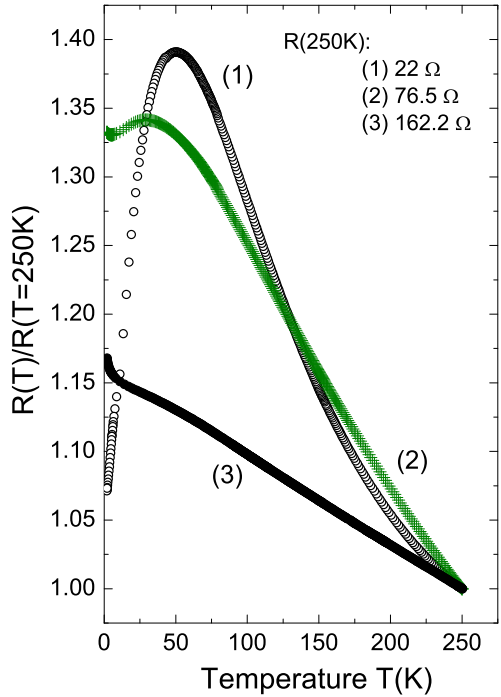}
\caption{Normalized resistance as a function of temperature for the
crystalline graphite flake in the: (1) as-prepared state, (2) after
the first ($5 \times 10^{11}~$cm$^{-2}$)  and (3) second ($1 \times
10^{12}~$cm$^{-2}$) Ga$^+$ irradiation.} \label{rtgra}
\end{figure}

Figure~\ref{rtgra} shows the temperature dependence of the graphite
sample in its three states, as prepared and after the two
irradiations. The temperature dependence in the as-prepared state has
a semiconducting behavior above 50~K and metallic below. The
semiconducting behavior is mainly due to the increase in carrier
concentration with temperature because most of the carriers are
thermally activated and the Fermi energy increases linearly with
temperature \cite{arn09}. The metallic behavior below 50~K is not
intrinsic of graphite but comes from internal interfaces between
crystalline regions parallel to the graphene layers but of slightly
different orientation\cite{bar08}. The mentioned interfaces originate
during the pyrolysis process \cite{interfaces}.

The first irradiation increased the resistance in all the temperature
range without changing strongly the relative change with temperature,
see Fig.~\ref{rtgra}. The metallic part was shifted to below 25~K and
its $T$-dependence gets weaker, suggesting that the irradiation also
affected the properties of the internal interface(s). After the
second irradiation the metallic region vanishes completely and the
resistance decreases rather linearly with $T$, see Fig.~\ref{rtgra}.
We note that the absolute value of the resistance is fairly
proportional to the used fluence. Between the virgin state and first
irradiation we have an increase in the resistance of a factor of 3.5.
Doubling the fluence we expect an increase of a factor $\sim 7$ of
the resistance of the as-received state, a factor in agreement with
the experimental observation.

As shown in \cite{arn09} the increase in resistance is due to the
decrease in the mean free path that overwhelms the increase in the
carrier density that these irradiations produce inside the graphite
structure. This behavior is related to difference weights the carrier
density $n$ and the mean free path $l$ have in the 2D resistivity,
i.e. $R \propto 1/(n^{1/2}l)$ in  contrast with the 3D resistivity
equation $R \propto 1/(nl)$. Note also that the second irradiation
produces already a vacancy density that implies a vacancy distance of
less than 2~nm in the graphene plane. At distances smaller than $\sim
3~$nm we do not expect a large increase in the carrier density with
further irradiation \cite{arn09}.

\begin{figure}[htb]%
\includegraphics[width=0.95\textwidth]{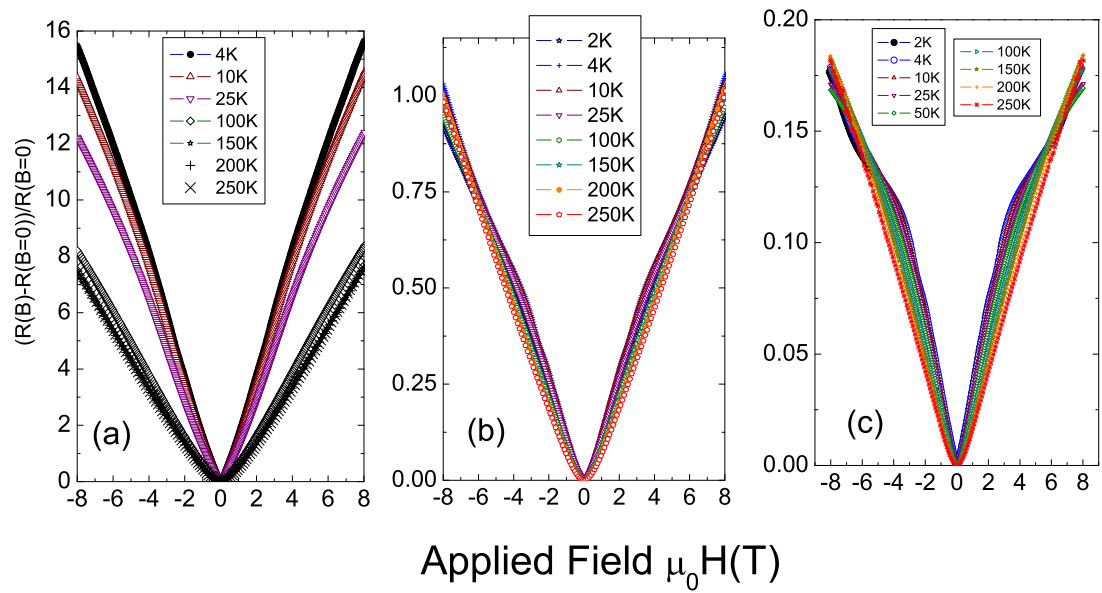}
\caption{Magnetoresistance as a function of field applied normal to
the graphene planes for the sample in its virgin state (a), after the
first (b) and second (c) irradiation. } \label{mrgra}
\end{figure}

Figure~\ref{mrgra} shows the magnetoresistance (MR) vs. applied field
at different constant temperatures for the graphite sample in the
virgin (a), first (b) and second (c) irradiated state. The MR in the
virgin state agrees with previous reports \cite{bar08}. Note that the
MR shows a quasi linear field behavior at low temperatures. Also
anomalous is the systematic increase of the MR below 100~K. This is
related to the decrease in the resistance with decreasing
temperature, see Fig.~\ref{rtgra}. The MR reaches a value of 16 at
8~T and 2~K. After the first irradiation the MR decreased a factor 16
and remains practically temperature independent. After the second
irradiation the MR decreased further a factor of 6 and shows a
similar temperature independence. The data reveal that the
Shubnikov-de Haas (SdH) oscillations increase their amplitude and
start to be measurable at lower fields after the first irradiation
\cite{arn09}. This behavior is related to the increase in the carrier
density due to the creation of defects, whereas the decrease in the
MR is due to the decrease in the mean free path.

\begin{figure}[htb]%
\includegraphics[width=0.7\textwidth]{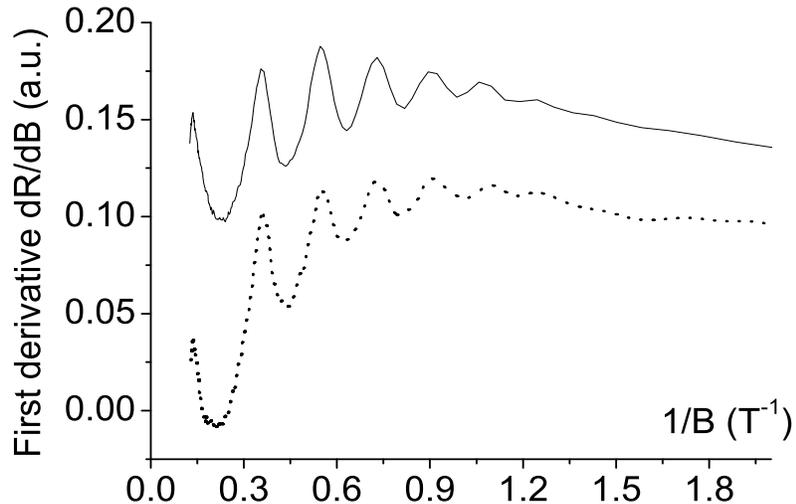}
\caption{First derivative of the resistance on field vs. inverse
field for the first (continuous line) and second (dotted line)
irradiation. The data have been multiplied by a factor in order to
show both derivatives in the same $y-$axis scale.} \label{sdh}
\end{figure}

\begin{figure}[htb]%
\includegraphics[width=0.6\textwidth]{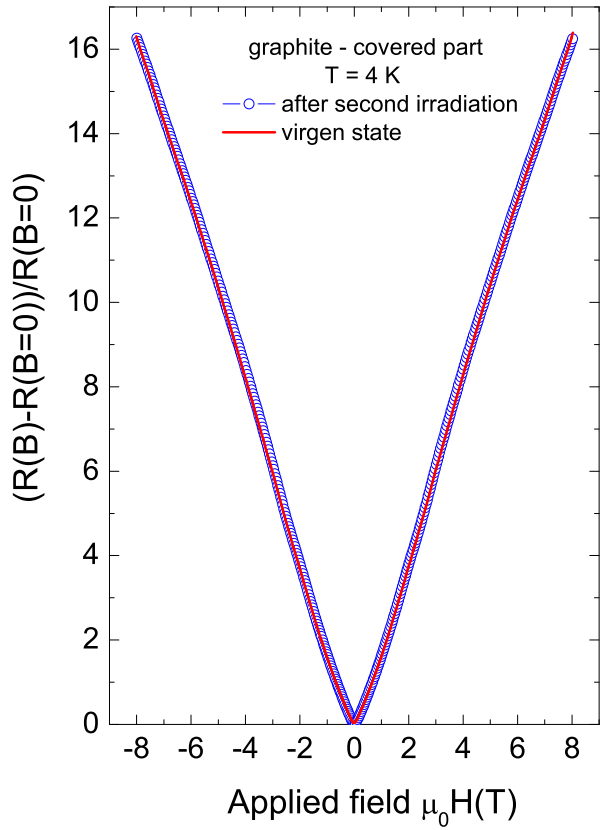}
\caption{Magnetoresistance of the upper covered part of the graphite
flake, see Fig.~\ref{samples_2}(a), at $T = 4~$K and in the virgin
state (continuous line) and after the second Ga$^+$ irradiation
($\circ$).} \label{MRcovered}
\end{figure}

Note that after the second irradiation the SdH oscillations do not
change qualitatively (their absolute amplitude changes due to the
large change of the MR), see Fig.~\ref{sdh}. After the second
irradiation we observe the first low-field oscillation at a similar
field and a similar oscillation period in reciprocal field as after
the first irradiation, see Fig.~\ref{sdh}. These results indicate
that no further change in the carrier density has been produced for a
vacancy distance less than 2~nm. Taking into account that 3~nm is of
the order of the range of modification of the electronic structure
produced by, e.g. a single vacancy \cite{rufi00}, then a saturation
of the carrier density is reached decreasing the vacancy distance
below $\sim 3~$nm but keeping the graphene structure. As expected,
the covered part of the sample does not show any change after the
first or second irradiation, as can be seen in Fig.~\ref{MRcovered}.

\begin{figure}[htb]%
\includegraphics[width=0.9\textwidth]{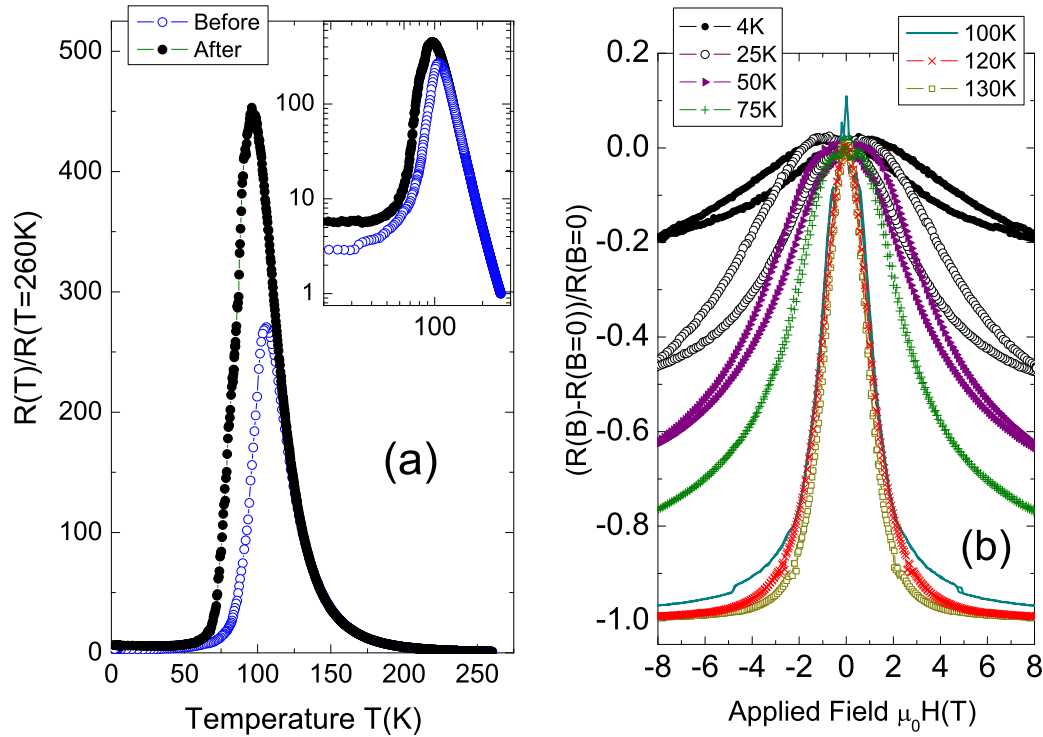}
\caption{(a) Temperature dependence of the normalized resistance
of the LCMO thin film before and after irradiation. The inset
shows the same data but in a double logarithmic scale. (b) The MR
vs. applied field for the as-prepared state. } \label{tlaca}
\end{figure}

\begin{figure}[htb]%
\includegraphics[width=0.9\textwidth]{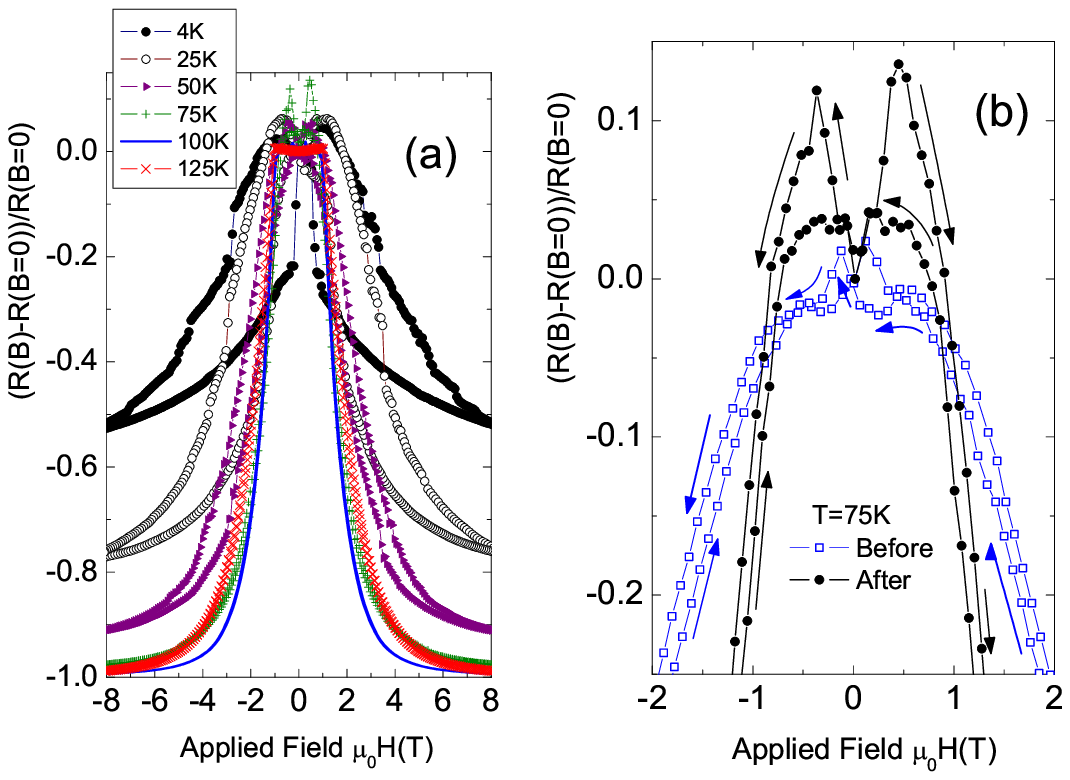}
\caption{(a) The MR of the LCMO film after irradiation. Compare
these results with those in Fig.~\protect\ref{tlaca}(b) and note
the induced changes in the MR by the irradiation. (b)
Magnetoresistance as a function of field at 75~K before and after
irradiation in a reduced field range to show the clear increase in
coercivity. This increase is observed at all temperatures in the
ferromagnetic state.} \label{hlaca}
\end{figure}
2.{\em LaCaMnO film}: The manganite sample undergoes a paramagnetic
insulator to ferromagnetic metal  transition leading to a sharp peak
in the resistance near its Curie temperature as shown in
Fig.~\ref{tlaca}(a). For our sample this peak is observed at $T_c =
106~$K. Interestingly, after irradiation the temperature dependence
does change only in the ferromagnetic state of the sample, which
shows now a $T_c = 95~$K, see Fig.~\ref{tlaca}(a). The measured MR of
this sample in the as-prepared state agrees with published literature
and shows hysteretic behavior in the ferromagnetic state whereas no
hysteresis in the paramagnetic state above $T_c$, see
Fig.~\ref{tlaca}(b).

The influence of ion irradiation on the magnetic and transport
properties of manganites were studied in the past but mainly at much
higher ion-energies, see e.g.
Refs.~\cite{str97,wol01,cha05,ram09,pal08}. In general at fluences
above $10^{12}~$cm$^{-2}$ the ion irradiation decreases the
metal-insulator transition temperature and the magnetic hysteresis
gets broader reflecting the increase of the pinning of the domain
walls by the induced defects. A similar behavior is observed in our
sample, see Fig.~\ref{hlaca}(a).  Figure~\ref{hlaca}(b) shows in
detail the MR between -2~T and 2~T at 75~K. The irradiation induces
an increase in the coercivity $H_c$, from 0.1~T to 0.42~T after
irradiation, defined at the maxima of the MR, as well as in the
overall hysteresis width.

These results indicate that care should be taken when changes in
the transport properties of ferromagnetic oxides are observed
after microstructuring the samples with FIB. The observed results
after patterning might not be due to the change of the sample
dimension but to the induced structural changes by the Ga$^+$
irradiation.

\begin{figure}[htb]%
\includegraphics[width=0.6\textwidth]{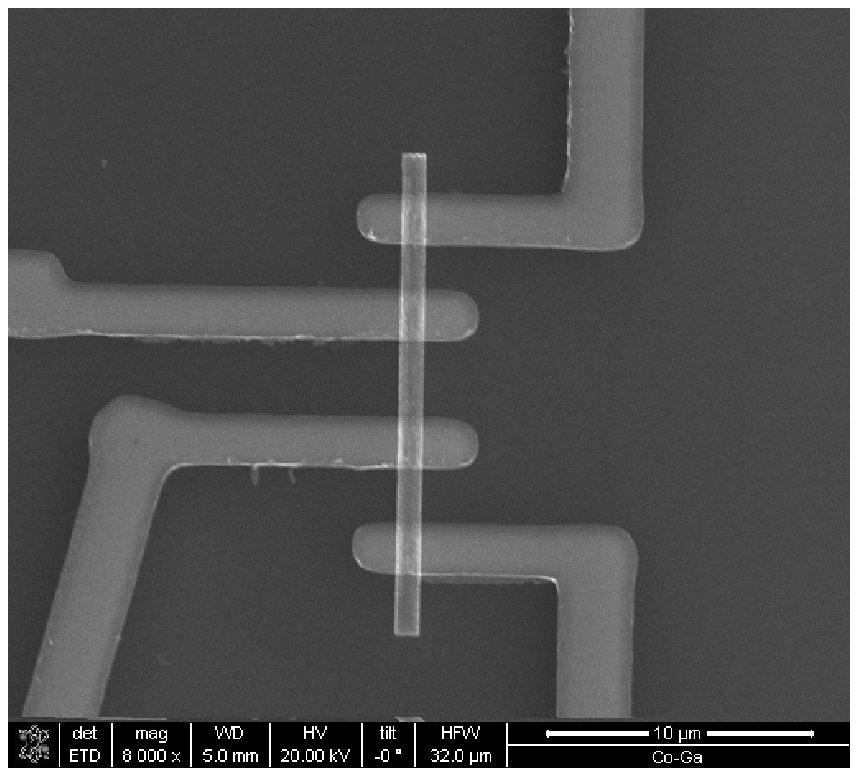}
\caption{Scanning electron microscope image of the Co sample ~\#2
with its four electrodes for resistance measurement. The thickness
of the wire was 22~nm.} \label{Co}
\end{figure}

\begin{figure}[htb]%
\includegraphics[width=0.9\textwidth]{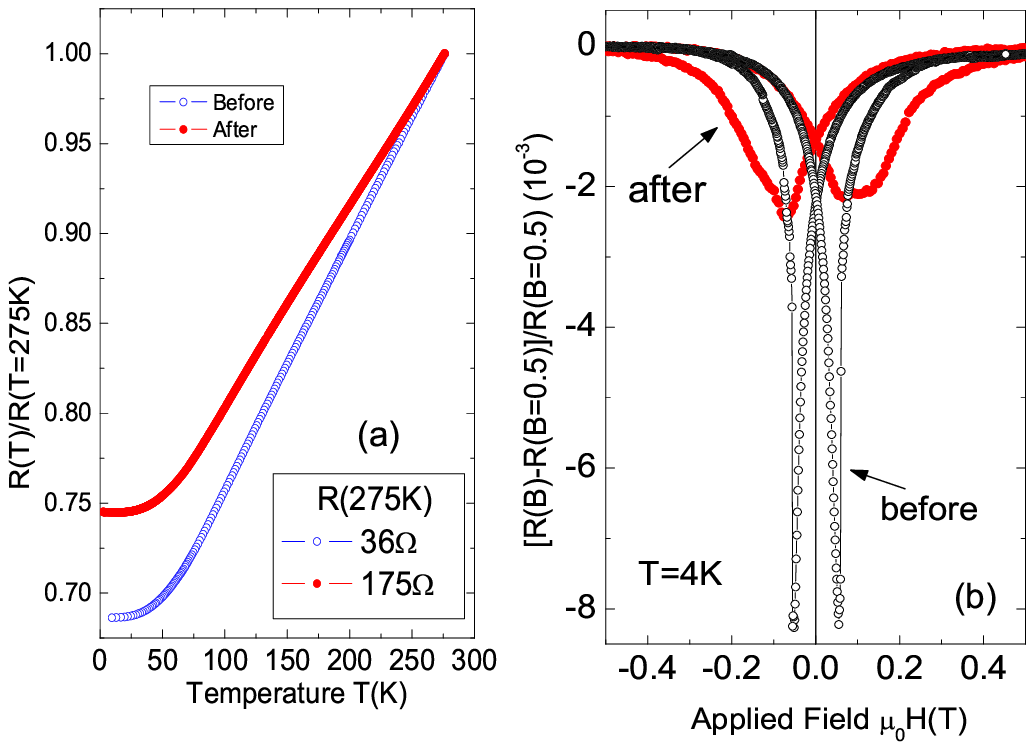}
\caption{(a) The normalized resistance of the Co~\#2 wire as a
function of temperature for the as-prepared and after irradiation
states. (b) MR of this sample for the two states at 4~K. The
Ga$^+$-irradiation fluence was $2.2 \times 10^{11}~$cm$^{-2}$.
Note the change in the MR at the coercivity field.} \label{Co-T-B}
\end{figure}

3.{\em Co film}: Figure~\ref{Co} shows a SEM picture of the
Co-sample \#2 with the four electrodes for the resistance
measurement. The resistance as a function of temperature and
magnetic field before and after a Ga$^+$ irradiation is shown in
Fig.~\ref{Co-T-B}(a) and (b). Due to the smaller thickness of this
sample the irradiation at 30~kV produces mainly defects instead of
doping since the Ga$^+$ ions stop beyond the sample thickness,
i.e. inside the substrate. The influence of the induced defects in
the Co sample can be clearly recognized by the increase in the
absolute value of the resistance. For example at 275~K the
resistance of the Co-sample~\#2 increases from $36~\Omega$ to
$175~\Omega$ flattening the temperature dependence, see
Fig.~\ref{Co-T-B}(a). The MR shown in Fig.~\ref{Co-T-B}(b)
indicates a clear change in the hysteresis indicating a change in
pinning of the domain walls. These results are qualitatively
similar to those obtained for the manganite shown above, see
Fig.~\ref{hlaca}.

\section{Conclusion}\label{con}

In conclusion our studies indicate clearly that care should be taken
with the change of the intrinsic properties of the materials when FIB
devices are used for patterning, cutting or just for deposit other
materials for contacts, for example. Our work demonstrates that
already the usual Ga$^+$ fluences needed for a precise alignment of
the Ga$^+$ beam before really using it, induce relevant changes in
the transport properties of the three different materials studied
here. Using a thin crystalline graphite flake we were able to
demonstrate also that covering the sample with a sufficiently thick
resist film one can avoid the irradiation damage completely. This
indicates that in principle one can use this technique to protect
certain parts and produce defined changes in other parts of the
sample of interest. This method might be used to  induce  changes in
the hysteretic properties of ferromagnetic micro and nano-structures
or in the electronic density in graphite or multigraphene in specific
parts of the sample, for example.

%\begin{acknowledgement}

\ack This work has been possible with the support of the DFG grant
DFG ES86/16-1. The authors S.D. and G.B. are supported by the
Graduate school BuildMona and the Collaborative Research Center (SFB
762) ``Functionality of Oxide Interfaces'', respectively.

%\end{acknowledgement}

\bibliographystyle{unsrt}
%\bibliography{D:/DATA/hopg/magnetic_carbon}
%\bibliography{D:/data/HOPG/magnetic_carbon}

\begin{thebibliography}{10}
\bibitem{interfaces}
As an example, one can take the process or graphitization of Kapton
  \cite{kapton,ina00}. Kapton is a polyimide, commercially available as thin
  films of different thickness. This polyimide has the advantage that through
  its carbonization at high temperatures and high pressures, it transforms into
  highly ordered graphite with properties similar to highly oriented pyrolytic
  graphite (HOPG) \cite{ina00}. In general, during the graphitization the
  crystalline regions grow through different steps at different temperatures
  and pressures where the release of different gases, like CO, CO$_2$, CH$_4$,
  H$_2$, N$_2$, occurs \cite{ina00}. Through the fabrication process these
  graphite single crystals within the carbon matrix do not grow with a
  thickness larger than $\sim 100~$nm \cite{bar08}, whereas the high degree of
  orientation remains in the ten micrometer size along the $a$ or $b$
  directions, i.e. parallel to the graphene planes \cite{ina00,gar08,bar08}. It
  appears plausible that structural defective regions, generated during the
  grow process and influenced by the release of gases, prevent a larger growth
  of the thickness of the crystalline regions, producing these planar
  interfaces.
\end{thebibliography}

\end{document}